\documentstyle[prb,epsfig,aps,multicol]{revtex}

\begin{document}
\title{The Dynamics of Structural Transitions in Sodium Chloride Clusters}
\author{Jonathan P.~K.~Doye and David J.~Wales}
\address{University Chemical Laboratory, Lensfield Road, Cambridge CB2 1EW, UK}
\date{\today}
\maketitle
\begin{abstract}
In recent experiments on sodium chloride clusters structural transitions 
between nanocrystals with different cuboidal shapes were detected.
Here we presents results for the thermodynamics and dynamics of one of these clusters, (NaCl)$_{35}$Cl$^-$.
As the time scales for the structural transitions can be much longer than those
accessible by conventional dynamics simulations, we use a master equation 
to describe the probability flow within a large sample of potential energy minima.
We characterize the processes contributing to probability flow between the different nanocrystals,
and obtain rate constants and activation energies for comparison with the experimental values.
\end{abstract}

\begin{multicols}{2}
\section{Introduction}

Understanding the relationship between the potential energy surface (PES), or
energy landscape, and the dynamics of a complex system is a major research 
effort in the chemical physics community.
One particular focus has been the dynamics of relaxation from an out-of 
equilibrium starting configuration down
the energy landscape to the state of lowest free energy, 
which is often also the global minimum of the PES.\cite{Doye96c}

The possible relaxation processes involved can be roughly divided into two types.
The first is relaxation from a high-energy disordered state 
to a low-energy ordered state, and examples include the folding of 
a protein from the denatured state to the native structure, 
and the formation of a crystal from the liquid.

The second kind of relaxation process, which is the focus of the work here, 
is relaxation from a low-energy, but metastable, state to
the most stable state. This second situation often arises from the first; 
the initial relaxation from a disordered state can lead to the population 
of a number of low-energy kinetically accessible configurations.
The time scales for this second relaxation process can be particularly long 
because of the large free energy barriers that can separate the states.

Some proteins provide an instance of this second type of relaxation.
Often, as well as a rapid direct path from the denatured state to the native structure,
there is also a slower path which passes through a low-energy kinetic trap.\cite{Miranker,Sosnick}
As this trapping is a potential problem for protein function, the cell has
developed its own biochemical machinery to circumvent it. 
For example, it has been suggested that the chaperonin, GroEL, aids protein folding 
by unfolding those protein molecules which get stuck in a trapped state.\cite{Shtilerman99} 

There are also a growing number of examples of this second type of relaxation involving clusters. 
For Lennard-Jones (LJ) clusters there are a number of sizes for which 
the global minimum is non-icosahedral.\cite{WalesD97}
For example, for LJ$_{38}$ the global minimum is a face-centred-cubic truncated octahedron,
but relaxation down the PES almost always leads to a low-energy icosahedral minimum. 
Similarly, for LJ$_{75}$ the icosahedral states act as a trap preventing relaxation to the Marks
decahedral\cite{Marks84} global minimum. A similar competition between face-centred-cubic or decahedral and
icosahedral structures has recently been observed for metal clusters.\cite{Cleveland98}

For these clusters unbiased global optimization is difficult because the icosahedral
states act as an effective trap. More generally, kinetic traps 
are one of the major problems for a global optimization algorithm.
Therefore, much research has focussed on decreasing the `life-time' of such traps. 
For example, some methods simulate a non-Boltzmann ensemble that involves increased fluctuations,
thus making barrier crossing more likely.\cite{Tsallis96,Andricoaei96} 
Other algorithms transform the energy landscape in a way that increases the temperature range 
where the global minimum is populated, thus allowing one to choose conditions 
where the free energy barriers relative to the thermal energy are lower.\cite{Doye98a,Doye98e}

Recently, clear examples of trapping associated with structural 
transitions have emerged from experiments on NaCl clusters.
These clusters have only one energetically favourable morphology:
the magic numbers that appear in mass spectra correspond to cuboidal
fragments of the bulk crystal (rocksalt) lattice,\cite{Campana81,Pflaum85,Twu90}
hence the term nanocrystals.
Indirect structural information comes from the experiments
of Jarrold and coworkers which probe the mobility of size-selected cluster ions.
For most (NaCl)$_N$Cl$^-$ with $N>30$,
multiple isomers were detected which were assigned as
nanocrystals with different cuboidal shapes.\cite{Dugourd97a}
The populations in the different isomers were not initially equilibrated, but
slowly evolved, allowing rates and activation energies for the
structural transitions between the nanocrystals to be obtained.\cite{Hudgins97a}

In a previous paper we identified the mechanisms of these structural
transitions by extensively searching the low-energy regions of the PES of one of these clusters, (NaCl)$_{35}$Cl$^-$, 
in order to obtain paths linking the different cuboidal morphologies.\cite{Doye99a}
The key process in these transitions is a highly cooperative rearrangement in which two parts 
of the nanocrystal slip past one another on a $\{110\}$ plane in a $\langle$1\={1}0$\rangle$ direction.

Here we continue our examination of the structural transitions by 
investigating the dynamics of (NaCl)$_{35}$Cl$^-$. Given the long time scales 
for which the clusters reside in metastable forms, it is not feasible to probe the
transitions with conventional dynamics simulations. 
Instead, we use a master equation that describes the probability flow between the minima on the PES. 
This method has the advantage that we can relate the dynamics to the topography of the PES .
In this paper we are particularly concerned with obtaining activation energies for the structural transitions: 
firstly, in order to compare with experiment, and secondly, to understand how the activation energy
for a process that involves a series of rearrangements and a large number of possible paths is related 
to the features of the energy landscape.
In section \ref{sect:methods} we outline our methods, and then in Section \ref{sect:results},
after a brief examination of the topography of the PES and the thermodynamics, we present our
results for the dynamics of the structural transitions.

\section{Methods}
\label{sect:methods}

\subsection{Potential}

The potential that we use to describe the sodium chloride clusters is 
the Tosi-Fumi parameterization of the Coulomb plus Born-Mayer form\cite{Tosi64}: 
\begin{displaymath}
E = \sum_{i<j}\left({q_iq_j\over r_{ij}} + A_{ij}e^{-r_{ij}/\rho}\right),
\end{displaymath}
where $q_i$ is the charge on ion $i$, $r_{ij}$ is the distance between ions
$i$ and $j$ and $A_{ij}$ and $\rho$ are parameters\cite{Tosi64}.
Although simple, this potential provides a reasonable description of the interactions.
For example, in a previous study we compared the global minima of 
(NaCl)$_{N}$Cl$^-$ ($N\le 35$) for this potential with those for a more complex 
potential derived by Welch {\it et al.\/} which also includes terms due 
to polarization.\cite{Welch75} 
Most of the global minima were the same for both potentials.\cite{Doye99a}
Given some of the other approximations we use in this study, 
the small advantages gained by using the Welch potential 
do not warrant the considerable additional computational expense. 

We should also note that a well-known problem associated with the
above family of potentials for the alkali halides is that they never 
predict the CsCl structure to be the most stable. 
This problem arises because the potentials do not allow the properties of an ion to
be dependent on the local ionic environment.\cite{Pyper94}
This deficiency should not greatly affect the relative energies of the 
low-lying (NaCl)$_{35}$Cl$^-$ minima because they all have the same rock-salt structure,
but it may affect the barriers for rearrangements where some ions experience 
a different local environment at the transition state. 

\subsection{Searching the potential energy surface}

The samples of 3518 minima and 4893 transition states that we use here were obtained 
in our previous study on the mechanisms of the structural transitions for (NaCl)$_{35}$Cl$^-$.\cite{Doye99a} 
This sampling was performed by repeatedly stepping across the PES from minimum to minimum 
via transition states, thus giving a connected set of stationary points. 
We biased this search to probe the low-energy regions of the PES either 
by using a Metropolis criterion\cite{Metropolis} 
to decide whether to accept a step to a new 
minimum\cite{Barkema96a,Doye97a,Mousseau97} 
or by systematically performing transition state searches from the 
lower energy minima in the sample.\cite{Tsai93a,Miller99a}
Thus, although our samples of stationary points 
only constitute a tiny fraction of the total number on the PES,
we have a good representation of the low-energy regions that 
are relevant to the structural transitions of (NaCl)$_{35}$Cl$^-$. 

\subsection{Thermodynamics}

We used two methods to probe the thermodynamics: first, conventional Monte Carlo simulations
and second, the superposition method. The latter is a technique to obtain
the thermodynamics of a system from a sample of minima.\cite{Wales93a}  It is based on the 
idea that all of configuration space can be divided up into the basins of attraction
surrounding each minimum.\cite{StillW84a} 
The density of states or partition function can then be written as a sum over all 
the minima on the PES, e.g. $Z=\sum_i Z_i$, where $Z_i$ is the partition function 
of minimum $i$. 

The limitations of the superposition method are that the $Z_i$ are not known exactly and that, for
all but the smallest systems, the total number of minima on the PES is too large for us
to characterize them all.
However, the harmonic expression for $Z_i$ leads to a reasonable description
of the thermodynamics.\cite{Wales93a} Furthermore, anharmonic forms are available which allow the
thermodynamics of larger clusters to be reproduced accurately.\cite{Doye95a}

The incompleteness of the sample can be overcome by weighting the contributions from the minima in 
a representative sample.\cite{Doye95a}
However, this approach is not necessary in the present study since we are interested in low temperature
behaviour where the number of thermodynamically relevant minima is still relatively small. 
Furthermore, in this temperature regime the superposition method has the advantage that it is unaffected by 
large free energy barriers between low-energy minima which can hinder the determination of 
equilibrium thermodynamic properties by conventional simulation.

Here we use the harmonic form of the superposition method, because we later use
the harmonic approximation to derive rate constants (reliable anharmonic expressions
for the rate constants are not so readily available). 
The partition function is then
\begin{equation}
Z=\sum_i {n_i e^{-\beta E_i}\over (\beta h \overline\nu_i)^\kappa},
\end{equation}
where $\beta=1/kT$, $E_i$ is the energy of minimum $i$, 
$\overline\nu_i$ is the geometric mean vibrational frequency of $i$,
$\kappa=3N-6$ is the number of vibrational degrees of freedom and
$n_i$ is the number of permutational isomers of $i$. 
$n_i$ is given by $2N!/h_i$, where $h_i$ is the order of the point group of $i$.
From this expression thermodynamic quantities such as the heat capacity, $C_v$, can 
be obtained by the application of standard thermodynamic formulae.\cite{Wales93a}  
The superposition method also allows us to examine the contributions of particular
regions of configuration space to the the thermodynamics. For example, the
probability that the system is in region $A$ is given by
\begin{equation}
\label{eq:pA}
p_A=\sum_{i\in A}{Z_i\over Z},
\end{equation}
where the sum runs over those minima that are part of region A.

\subsection{Dynamics}

The master equation approach\cite{vanK} is increasingly being used to describe
the inter-minimum dynamics on a multi-dimensional PES with applications to, for example,
clusters,\cite{BerryK95,Kunz95,Ball98b,Miller99b} 
glasses,\cite{Angelani98,Angelani99} proteins\cite{Czerminski90,Cieplak98a,Westerberg99} 
and idealized model landscapes.\cite{Doye96c} 
The master equation is defined in terms of ${\bf P(t)}=\{P_i(t)\}$, 
the vector whose components are the ensemble-average probabilities that the system is associated 
with each minimum at time $t$:
\begin{equation}
\frac{dP_i(t)}{dt}=\sum_{j\ne i}^{n_{\rm min}}[k_{ij}P_j(t)-k_{ji}P_i(t)],
\label{eq:dpdt}
\end{equation}
where $k_{ij}$ is the rate constant for transitions from minimum
$j$ to minimum $i$.
Defining the matrix
\begin{equation}
W_{ij}=k_{ij}-\delta_{ij}\sum_{m=1}^{n_{\rm min}} k_{mi}
\end{equation}
allows Equation (\ref{eq:dpdt}) to be written in matrix form: 
$d{\bf P}(t)/dt = {\bf WP}(t)$.

If the transition matrix $\bf W$ cannot be decomposed into block form, it has a single 
zero eigenvalue whose corresponding eigenvector is the equilibrium probability
distribution, ${\bf P}^{eq}$. 
As a physically reasonable definition for the rate constants must obey
detailed balance at equilibrium, i.e.\ $W_{ij}P^{\rm eq}_j = W_{ji}P^{\rm eq}_i$, the
solution of the master equation can be expanded in terms of a complete set of 
eigenfunctions of the symmetric matrix, ${\bf \tilde W}$, defined by 
$\tilde W_{ij}=(P^{\rm eq}_j/P^{\rm eq}_i)^{1/2} W_{ij}$. The solution is
\begin{equation}
P_i(t)=\sqrt{P^{\rm eq}_i}\sum_{j=1}^{n_{\rm min}}{\tilde u}_i^{(j)} e^{\lambda_j t}
\left[\sum_{m=1}^{n_{\rm min}}{\tilde u}_m^{(j)}\frac{P_m(0)}{\sqrt{P^{\rm eq}_m}}\right],
\label{eq:msolution}
\end{equation}
where ${\tilde u}_i^{(j)}$ is component $i$ of the $j^{\rm th}$ eigenvector 
of ${\bf \tilde W}$ and $\lambda_j$ is the $j^{\rm th}$ eigenvalue.

The eigenvalues of  $\bf W$ and ${\bf \tilde W}$ are identical and the eigenvectors are
related by $u_i^{(j)}={\tilde u}_i^{(j)}\sqrt{P_i^{eq}}$. 
Except for the zero eigenvalue, all $\lambda_j$ are negative. 
Therefore, as $t\rightarrow\infty$ the contribution of these modes decays
exponentially to zero and ${\bf P}\rightarrow {\bf P}^{eq}$.

To apply Equation (\ref{eq:msolution}) we must first diagonalize ${\bf \tilde W}$. 
The computer time required for this procedure scales as the cube of the size 
of the matrix and the memory requirements scale as the square. 
Therefore, it is advantageous for the matrix ${\bf \tilde W}$ to be as small as possible.
For this reason we recursively removed those minima that are only connected to one other minimum;
these `dead-end' minima do not contribute directly to the probability 
flow between different regions of the PES.
After pruning our samples have 1624 minima and 2639 transition states.
To test the effect of this pruning, we performed some 
calculations using both the full and the pruned samples.
The effect on the dynamics of the structural transitions was negligible.

As the temperature of a system is decreased the spread of eigenvalues can increase rapidly.
When the ratio of the largest to smallest eigenvalues is of the order of 
the machine precision of the computer, the accuracy of the extreme eigenvalues 
can become degraded by rounding errors.
We encountered these problems below 275K.
Without pruning the samples these numerical difficulties are more pronounced.\cite{Miller99b}

We model the rate constants, which are needed as input to Equation (\ref{eq:dpdt}),
using RRKM theory\cite{Forst} in the harmonic approximation. 
Therefore, the rate constant for a transition from minimum $i$ to minimum $j$ via
a particular transition state (denoted by $\dagger$) is given by
\begin{equation}
k^\dagger_{ij}(T)=\frac{h_i}{h_{ij}^\dagger}
\frac{\bar\nu_i^\kappa}{\bar\nu_{ij}^{\dagger(\kappa-1)}}
\exp(-(E^\dagger_{ij}-E_i)/kT).
\end{equation}

\section{Results}
\label{sect:results}

\subsection{Topography of the (NaCl)$_{35}$Cl$^-$ PES}

\begin{center}
\begin{figure}
\vglue-1.6cm
\epsfig{figure=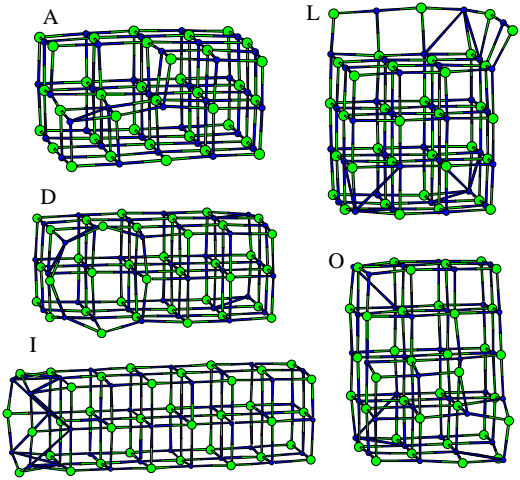,width=8.2cm}
\vglue-2.5cm
\begin{minipage}{8.5cm}
\caption{\label{fig:minima}
The lowest-energy $5\times 5\times 3$ (A), $6\times 4\times 3$ (D) 
and $8\times 3\times 3$ (I) nanocrystals and
the two lowest-energy $5\times 4\times 4$ nanocrystals (L and O).
The sodium ions are represented by the smaller, darker circles and
the chloride ions by the larger, lighter circles.
The letter gives the energetic rank of the minimum when labelled alphabetically.}
\end{minipage}
\end{figure}
\end{center}

In our previous study of (NaCl)$_{35}$Cl$^-$ we found that the low-energy minima all had
rock-salt structures. 
The different minima have 
four basic shapes: an incomplete $5\times 5\times 3$ cuboid,
a $6\times 4\times 3$ cuboid with a single vacancy, a $8\times 3\times 3$ cuboid with
a single vacancy and an incomplete $5\times 4\times 4$ cuboid. 
The lowest-energy minimum for each of these forms is shown in Figure \ref{fig:minima}.

In the experiments on (NaCl)$_{35}$Cl$^-$ by Jarrold and coworkers the three peaks 
that were resolved in the arrival time distribution were assigned on 
the basis of calculated mobilities as $5\times 5\times 3$, 
$5\times 4\times 4$ and $8\times 3\times 3$ nanocrystals.\cite{Dugourd97a,Hudgins97a}
However, when the $6\times 4\times 3$ nanocrystal is also considered, 
better agreement between the calculated and observed mobilities
can be obtained by assigning the three experimental peaks to the $6\times 4\times 3$,
$5\times 5\times 3$ and $8\times 3\times 3$ nanocrystals in order of increasing 
drift time.\cite{Doye99a,Alexprivate} This reassignment is also in better agreement
with the thermodynamics since the clusters convert to (what is now assigned as)
the $5\times 5\times 3$ nanocrystal as time progresses,\cite{Hudgins97a} indicating that this
structure has the lowest free energy. In our calculations a $5\times 5\times 3$ 
isomer is the global potential energy minimum, and the $5\times 5\times 3$ nanocrystal is always more 
stable than the other nanocrystals (See Section \ref{sect:therm}).

Disconnectivity graphs provide a way of visualizing an
energy landscape that is particularly useful for obtaining insight 
into dynamics, and have previously been applied to 
a number of protein models\cite{Czerminski90,Becker97,Levy98a,Miller99c,Garstecki99} 
and clusters.\cite{Miller99a,WalesMW98,Doye99c,Doye99f}
The graphs are constructed by performing a `superbasin' analysis at 
a series of energies. This analysis involves grouping minima into
disjoint sets, called superbasins, whose members are connected by pathways that
never exceed the specified energy. At each energy a superbasin is represented by
a node, and lines join nodes in one level to their daughter nodes in the 
level below. Every line terminates at a local minimum. 
The graphs therefore present a visual representation of the hierarchy of barriers 
between minima.

\begin{center}
\begin{figure}
\epsfig{figure=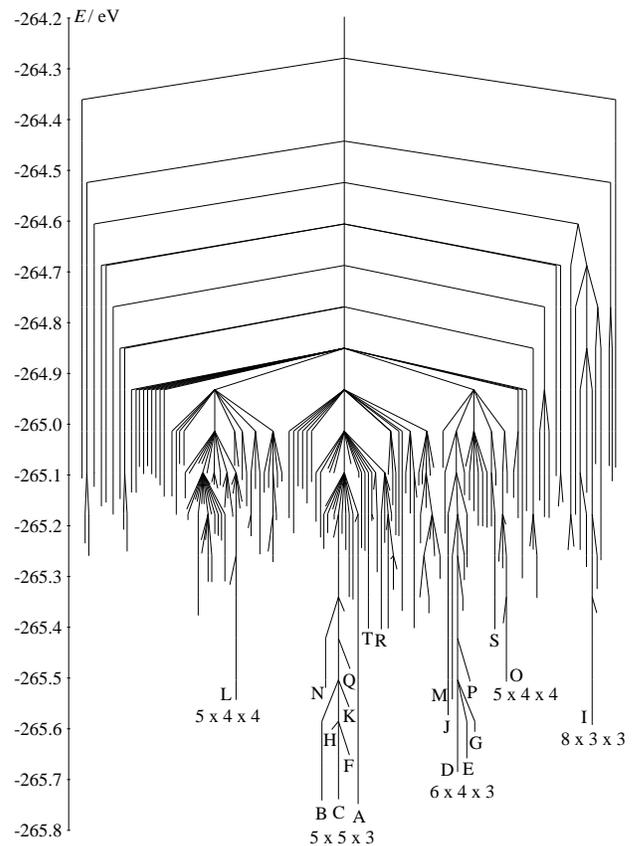,width=8.2cm}
\vglue0.1cm
\begin{minipage}{8.5cm}
\caption{\label{fig:disconnect} 
Disconnectivity graph for (NaCl)$_{35}$Cl$^-$. 
Only branches leading to the lowest 200 minima are shown.
The branches for the twenty lowest-energy minima are labelled 
alphabetically in order of their energy.
Some of the funnels and sub-funnels have also been labelled by their
associated cuboidal shape.
}
\end{minipage}
\end{figure}
\end{center}

The disconnectivity graph for (NaCl)$_{35}$Cl$^-$ is shown in Figure \ref{fig:disconnect}.
The barriers between minima with the same cuboidal form are generally lower than those between 
minima that have different shapes. Therefore, the disconnectivity graph 
splits into funnels corresponding to each cuboidal morphology.
(A funnel is a set of downhill pathways that converge on a single low-energy minimum or 
a set of closely-related low-energy minima.\cite{Leopold,Bryngelson95} 
In a disconnectivity graph an ideal funnel is 
represented by a single tall stem with lines branching directly from it, 
indicating the progressive exclusion of minima as the energy is decreased.\cite{WalesMW98})
The separation is least clear for the $5\times 4\times 4$ minima because of the large
number of different ways that the nine vacant sites can be arranged. 
For example, these vacancies are organized very differently in 
the two lowest-energy $5\times 4\times 4$ minima (Figure \ref{fig:minima}),
and in fact the barrier between minimum O and the low-energy $6\times 4\times 3$ isomers
is lower than the barrier between O and minimum L.
Therefore, the minima associated with O form a sub-funnel that splits off from 
the $6\times 4\times 3$ funnel, rather than being directly connected to the main 
$5\times 4\times 4$ funnel.

The disconnectivity graph shows that the barriers between the 
$5\times 5\times 3$, $6\times 4\times 3$ and $5\times 4\times 4$ 
nanocrystals are of similar magnitude, while the $8\times 3\times 3$ minima are separated 
from the rest by a considerably larger barrier.
The values of some of the barrier heights are given in Table \ref{table:barriers}.

The disconnectivity graph is also helpful for interpreting the (NaCl)$_{35}$Cl$^-$ dynamics observed
in experiments.\cite{Dugourd97a,Hudgins97a} In the formation process 
it is likely that a high-energy configuration is initially generated. 
The cluster then relaxes to one of the low-energy nanocrystals.
Simulations for potassium chloride clusters indicate that this relaxation is particularly
rapid for alkali halides because of the low barriers (relative to the energy difference 
between the minima) for downhill pathways.\cite{Rose93b,Ball96}
However, there is a separation of time scales between this initial relaxation and
the conversion of the metastable nanocrystals to the one with the lowest free energy.
the large barriers between the different cuboids make them efficient traps.

\begin{center}
\begin{figure}
\epsfig{figure=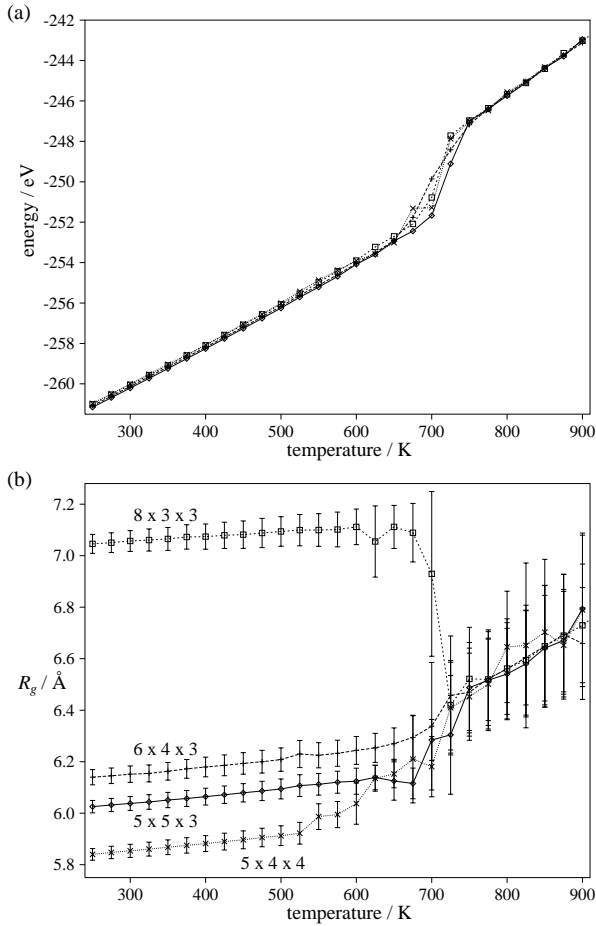,width=8.2cm}
\vglue0.1cm
\begin{minipage}{8.5cm}
\caption{\label{fig:MC} 
The temperature dependence of (a) the total energy and 
(b) the radius of gyration, $R_g$, for four series of MC runs starting in the 
lowest-energy minima of the $5\times 5\times 3$ (diamonds), $6\times 4\times 3$ (plus signs), 
$8\times 3\times 3$ (squares) and $5\times 4\times 4$ (crosses) nanocrystals.
Each point is the average value in a $10^6$ cycle Monte Carlo run, where 
each run was initiated from the final geometry in the previous lower temperature run.
The error bars in (b) represent the standard deviation of the $R_g$ probability distributions.
}
\end{minipage}
\end{figure}
\end{center}

\subsection{Thermodynamics of (NaCl)$_{35}$Cl$^-$}
\label{sect:therm}

\begin{center}
\begin{figure}
\epsfig{figure=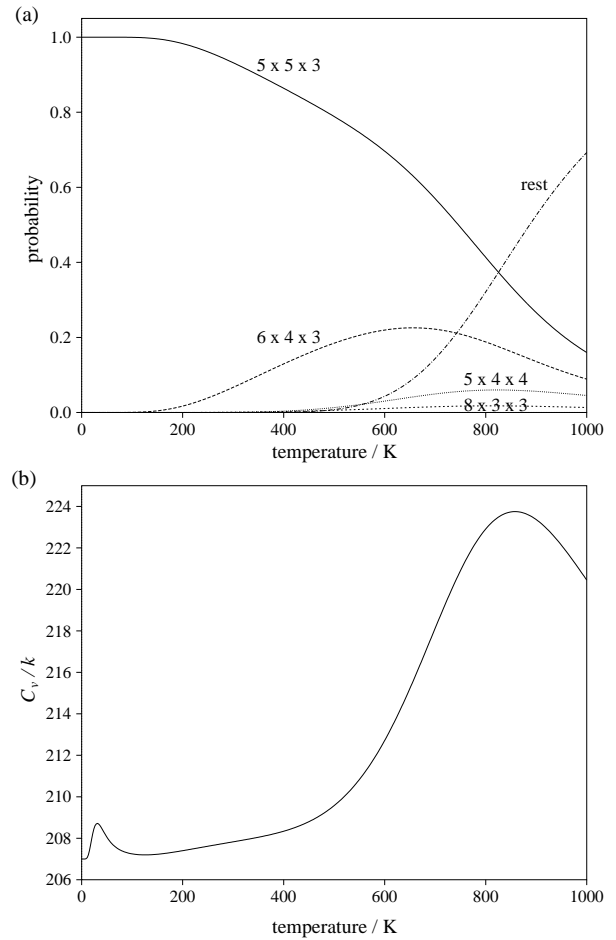,width=8.2cm}
\vglue0.1cm
\begin{minipage}{8.5cm}
\caption{\label{fig:HSM} Thermodynamic properties of (NaCl)$_{35}$Cl$^-$ computed using the 
harmonic superposition method with the pruned sample of minima.
(a) Equilibrium occupation probabilities of the nanocrystals, as labelled.
(b) Heat capacity.
}
\end{minipage}
\end{figure}
\end{center}

Some thermodynamic properties of (NaCl)$_{35}$Cl$^-$ are shown in Figures \ref{fig:MC} and \ref{fig:HSM}.
The caloric curve shows a feature at $\sim$700K which indicates melting (Figure \ref{fig:MC}a);
the melting temperature is depressed relative to the bulk due to the cluster's finite size.
The effects of the barriers between nanocrystals 
are apparent in the MC simulations. 
The radius of gyration, $R_g$, provides a means of differentiating the nanocrystals.
It can be seen from the plot of $R_g$ for simulations started in the lowest-energy minima 
of the four cuboidal forms that each simulation is stuck in the starting 
structure (Figure \ref{fig:MC}b) up to temperatures close to melting,
implying that there are large free energy barriers between the nanocrystals. 

These free energy barriers prevent an easy determination of the relative stabilities of 
the different nanocrystals by conventional simulations. Therefore, we use the 
superposition method to examine this question. First we assign the fifty lowest-energy minima
to one of the cuboidal forms by visual inspection of each structure. 
Then, using these sets as definitions of the nanocrystals in Equation (\ref{eq:pA}), 
we calculate the equilibrium probabilities of the cluster 
being in the different cuboidal morphologies as a function of temperature.
It can be seen from Figure \ref{fig:HSM}a that the $5\times 5\times 3$ nanocrystal is most 
stable up until melting. The $6\times 4\times 3$ nanocrystal also has a significant
probability of being occupied. However, the probabilities for the $8\times 3\times 3$
and $5\times 4\times 4$ nanocrystals are always small. 
The onset of the melting transition is indicated by the rise of 
$p_{\rm rest}$ in Figure \ref{fig:HSM}a and by the peak in the heat capacity (Figure \ref{fig:HSM}b). 
However, this transition is much too broad and the heat capacity peak occurs at 
too high a temperature because the incompleteness of our sample of minima
leads to an underestimation of the partition function for the liquid-like minima.
Given these expected failings of the superposition method at high temperature
when the partition functions of the minima are not reweighted, 
we restrict our dynamics calculations to temperatures below 600K.

Although the probabilities of being in the different morphologies 
show little variation at low temperature, there are significant changes in the occupation probabilities 
of specific minima. For example, the small low temperature peak in the heat capacity is 
a result of a redistribution of probability amongst the low-energy $5\times 5\times 3$
minima; the third lowest-energy minimum becomes most populated.
It is also interesting to note that the second lowest-energy $5\times 4\times 4$ minimum (O) becomes more
stable than the lowest-energy $5\times 4\times 4$ minimum (L) for temperatures above approximately 220K. 
Both these changes are driven by differences in vibrational entropy.

\subsection{Dynamics of (NaCl)$_{35}$Cl$^-$}

\begin{center}
\begin{figure}
\epsfig{figure=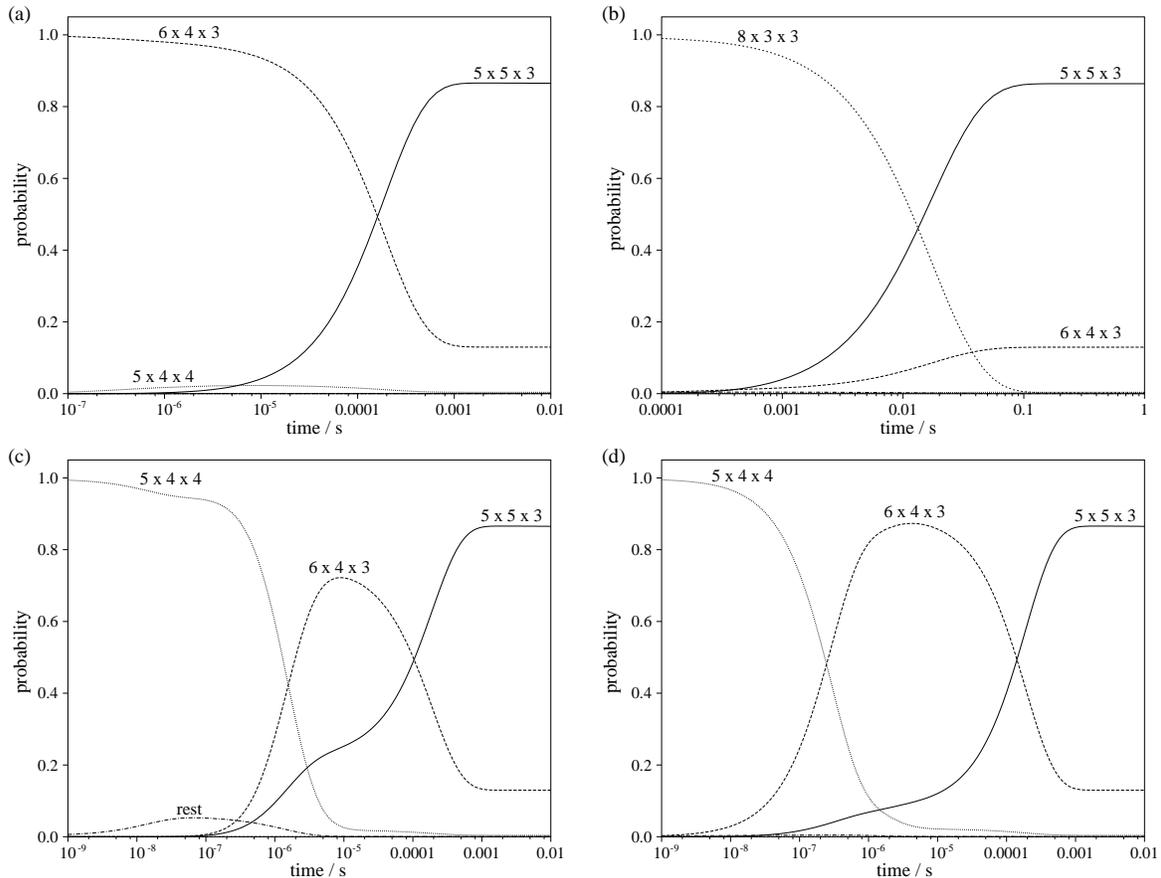,width=15.7cm}
\vglue0.1cm
\begin{minipage}{17.6cm}
\caption{\label{fig:flow} 
Time evolution of the occupation probabilities of the different 
nanocrystals at $T$=400K when the cluster is initially in minimum (a) D,
(b) I, (c) L and (d) O.
}
\end{minipage}
\end{figure}
\end{center}

Some examples of the interfunnel dynamics that we find on solution of 
the master equation are depicted in Figure \ref{fig:flow}. 
The time scales involved are much longer than those accessible 
by conventional simulations.

The dynamics of relaxation to equilibrium depend significantly on the 
starting configuration. When the lowest-energy $6\times 4\times 3$ minimum is the initial
configuration there is a small transient population in $5\times 4\times 4$ minima before the system
adopts a $5\times 5\times 3$ structure. This is consistent with the lowest-energy pathway that was 
found between the two nanocrystals; it passes through
some intermediate $5\times 4\times 4$ minima.\cite{Doye99a} 

\end{multicols}
\begin{multicols}{2}

The relaxation to equilibrium is much slower when the cluster starts from the 
lowest-energy $8\times 3\times 3$ minimum. This is a result of the large barrier to escape from this 
funnel (Figure \ref{fig:disconnect}).
The probability flow out of the $8\times 3\times 3$ funnel leads to a simultaneous rise in the 
occupation probabilities of both the $5\times 5\times 3$ and $6\times 4\times 3$ nanocrystals towards their
equilibrium values, even though the lowest-energy pathway out of the $8\times 3\times 3$ funnel 
directly connects it to the low-energy $6\times 4\times 3$ minima.  
This occurs because 
the time scale for interconversion of these latter two nanocrystals 
is much shorter than that for escape from the $8\times 3\times 3$ funnel. 
However, we do find evidence that the cluster first passes through the $6\times 4\times 3$ minima
if we examine the probabilities on a log-scale. 
At times shorter than that required for local equilibrium between the $5\times 5\times 3$ 
and $6\times 4\times 3$ minima the occupation probability for the $6\times 4\times 3$ minima is larger.

As the two lowest-energy $5\times 4\times 4$ minima are well-separated in configuration space we considered
relaxation from both these minima. 
In both cases, there is a large probability flow into the $6\times 4\times 3$ minima, which is then transferred 
to the $5\times 5\times 3$ funnel on the same time scale as when initiated in the $6\times 4\times 3$ funnel.
However, the time scale for the build-up of population in the $6\times 4\times 3$ minima 
depends on the initial configuration. 
Probability flows more rapidly and directly into the $6\times 4\times 3$ minima when initiated from minimum O,
reflecting the low barriers between these minima (Figure \ref{fig:disconnect}). 
For relaxation from minimum L there are two active pathways, 
leading to an increase in the population of both the $5\times 5\times 3$ and the $6\times 4\times 3$ minima.
The direct path into the $5\times 5\times 3$ funnel has the lower barrier (Table \ref{table:barriers}) 
but is long (96.7\AA),
and so has a smaller rate than the path into the $6\times 4\times 3$ funnel, 
which has a slightly higher barrier (by 0.05$\,$eV).
The small shoulder in the occupation probability of the $5\times 5\times 3$ minima occurs
in the time range when the occupation probability of the $5\times 4\times 4$ minima has reached a value
close to zero (thus reducing the contribution from the direct path) and when the probability 
flow out of the $5\times 4\times 4$ minima is only just beginning.

The combination of our thermodynamics and dynamics results for (NaCl)$_{35}$Cl$^-$ enable us to explain
why a peak associated with the $5\times 4\times 4$ cuboids was not observed experimentally. 
The $5\times 4\times 4$ minima 
have a shorter lifetime than the other cuboidal forms and have a low equilibrium occupation probability.

Another way to analyse the dynamics is to examine 
how local equilibration progresses 
towards the point where global equilibrium has been obtained. 
To accomplish this we define two minima to be in local equilibrium at the time when 
\begin{equation}
\frac{\left|P_i(t)P^{\rm eq}_j-P_j(t)P^{\rm eq}_i\right|}
{\sqrt{P_i(t)P_j(t)P^{\rm eq}_iP^{\rm eq}_j}} \leq\epsilon
\label{eq:eqtime}
\end{equation}
is obeyed for all later times. In the present work we set $\epsilon=0.01$, i.e.\ 
the two minima are within 1\% of equilibrium. Using this definition we can 
construct equilibration graphs, in which nodes occur when two (groups of) minima
come into local equilibrium.\cite{Sibani}

\begin{center}
\begin{figure}
\epsfig{figure=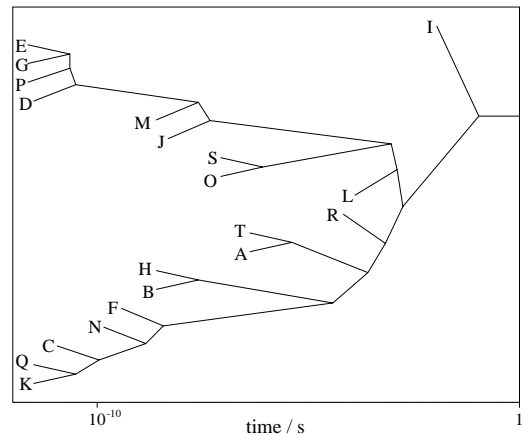,width=8.2cm}
\vglue0.1cm
\begin{minipage}{8.5cm}
\caption{\label{fig:equil} 
Equlibration graph showing how the system progresses towards equilibrium 
at $T$=400K when the cluster is initially in minimum D. 
The lines join when minima come into equilibrium with each other. 
The vertical scale and the horizontal position of the ends 
of the lines are chosen for clarity. 
The ends are labelled by the letter for the corresponding minimum.
}
\end{minipage}
\end{figure}
\end{center}

We show an example of an equilibration graph in Figure \ref{fig:equil}.
Equilibration first occurs at the bottom of the $6\times 4\times 3$ and $5\times 5\times 3$ funnels between
those minima that are connected by low barrier paths, then progresses 
to minima with the same cuboidal shape but which are separated by 
larger barriers, and finally occurs between the funnels. The order of interfunnel 
equilibrium agrees with the time scales that we observe in the time evolution 
of the occupation probabilities of the nanocrystals (Figure \ref{fig:flow}). 
Minimum O, then minimum L, come into equilibrium with the $6\times 4\times 3$ funnel. 
Then, the $5\times 5\times 3$ and $6\times 4\times 3$ funnels reach local equilibrium. 
Finally, the $8\times 3\times 3$ funnel reaches equilibrium with the rest of the PES.
As one of the major determinants of the time scale required for local equilibrium is the height
of the barriers between minima, it is unsurprising that the equilibration graph reflects the 
structure of the disconnectivity graph (Figure \ref{fig:disconnect}).

In the experiments on (NaCl)$_{35}$Cl$^-$ rate constants and activation energies were obtained
for the conversion of the $6\times 4\times 3$ and $8\times 3\times 3$ nanocrystals into the $5\times 5\times 3$ nanocrystal.\cite{Hudgins97a} 
It would, therefore, be useful if we could extract rate constants for the different 
interfunnel processes from the master equation dynamics. 

For a two-state system, where $A \rightleftharpoons B$ and $k_+$ and $k_-$ are 
forward and reverse rate constants, respectively, it can be shown that
\begin{equation}
\ln\left[\frac{P_A(t)-P^{\rm eq}_A}
{P_ A(0)-P^{\rm eq}_ A}\right]
=-(k_+ + k_-)t,
\label{eq:twostate}
\end{equation}
and the equivalent expression for B are obeyed.\cite{Miller99b} 
This is a standard result for a first-order reaction.
This expression will also hold for the rate of passage between two funnels in 
our multi-state system if the interfunnel dynamics are the only processes affecting the 
occupation probabilities of the relevant funnels, and if the interfunnel dynamics
cause the occupation probabilities of the two funnels to converge to their equilibrium values. 

\begin{center}
\begin{figure}
\epsfig{figure=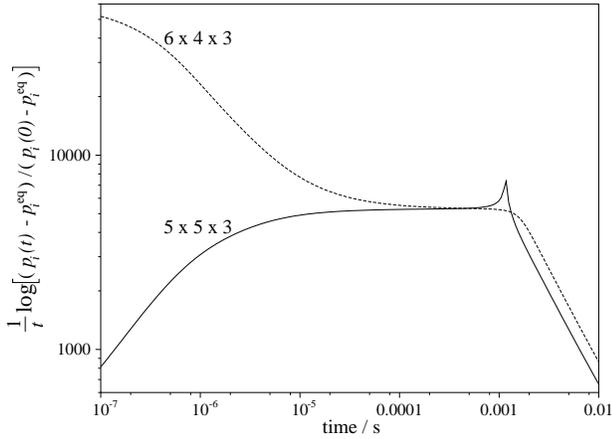,width=8.2cm}
\vglue0.1cm
\begin{minipage}{8.5cm}
\caption{\label{fig:first} 
By following $(1/t) \log [ (p_i(t)-p_i^{\rm eq})/(p_i(0)-p_i^{\rm eq})]$ at $T$=400K 
when minimum D is the initial configuration, this graph tests the applicability of 
Equation (\ref{eq:twostate}) to the interconversion of the $5\times 5\times 3$ and $6\times 4\times 3$ nanocrystals.
}
\end{minipage}
\end{figure}
\end{center}

In Figure \ref{fig:first} we test the above expression by applying it to 
the interconversion of $6\times 4\times 3$ and $5\times 5\times 3$ nanocrystals. 
The two lines in the graph converge to the same plateau value, before both falling off beyond 0.001s.
This plateau corresponds to the time range for which the interfunnel passage dominates
the evolution of the probabilities for the two funnels. 
At shorter times, when the occupation probabilities for the two funnels 
are still close to their initial values,
there are many other contributing processes.
At longer times the probabilities are both very close to their equilibrium values, 
and the slower equilibration with the $8\times 3\times 3$ funnel dominates the probability evolution.
From the plateau in Figure \ref{fig:first} we obtain $k^+ + k^-=5320\,{\rm s}^{-1}$. 
The individual rate constants can be obtained by using the detailed balance relation:
$k^+ P_A^{\rm eq}=k^- P_B^{\rm eq}$.

The application of Equation (\ref{eq:twostate}) to escape from the $8\times 3\times 3$ funnel also 
leads to a range of $t$ where there is a well-defined plateau. 
However, this approach works less well for the interconversion of 
$5\times 4\times 4$ and $6\times 4\times 3$ minima (Figures \ref{fig:flow}a and b). 
This is because the assumption that no other processes contribute to the
probability evolution of the two funnels is obeyed less well,
and because the occupation probabilities do not converge to their 
equilibrium values, but near to the equilibrium values that would 
obtain if the $5\times 5\times 3$ minima were excluded.
Nevertheless, approximate values of $k^+$ and $k^-$ can be obtained.

Diagonalization of the matrix ${\bf \tilde W}$ produces a set of eigenvalues 
that give the time scales for a set of characteristic probability flows. 
The dynamical processes to which the eigenvalues correspond can be 
identified by examining the eigenvectors.
Flow occurs between those minima for which the corresponding components
of the eigenvector have opposite sign (Equation (\ref{eq:msolution})). 
This observation forms the basis for Kunz and Berry's net-flow index which quantifies 
the contribution of an eigenvector $i$ to flow out of a funnel $A$.\cite{BerryK95,Kunz95} 
The index is defined by 
\begin{equation}
f^{\rm A}_i=\sum_{j\in{\rm A}}\tilde u^{(i)}_j \sqrt{P^{\rm eq}_j}.
\label{eq:flow}
\end{equation}
The index allows the interfunnel modes to be identified; 
the values of $f^A$ and $f^B$ for these modes will be large and of opposite sign.
For example, at $T$=400K the mode with the most $5\times 5\times 3\rightarrow 6\times 4\times 3$ character has
$f^{5\times 5\times 3}=-0.339$, $f^{6\times 4\times 3}=0.331$ and $\lambda=5275\,{\rm s}^{-1}$. 
The eigenvalue is in good agreement with the sum of interfunnel rates obtained 
using Equation (\ref{eq:twostate}). 

The extraction of the interfunnel rate in this manner is hindered by the fact that 
the eigenvalues of ${\bf \tilde W}$ cannot cross as a function of temperature. 
Instead, there are avoided crossings and mixing of modes. 
For example, the small difference between the two values for the sum of the
$5\times 5\times 3\leftrightarrow 6\times 4\times 3$ interfunnel
rate constants that we obtained above is probably due to mixing. 
The eigenvector with which the interfunnel mode mixes 
gains some $5\times 5\times 3\rightarrow 6\times 4\times 3$ character;
it has $f^{5\times 5\times 3}=-0.014$, $f^{6\times 4\times 3}=0.017$ and $\lambda=6388{\rm s}^{-1}$.

\begin{center}
\begin{figure}
\epsfig{figure=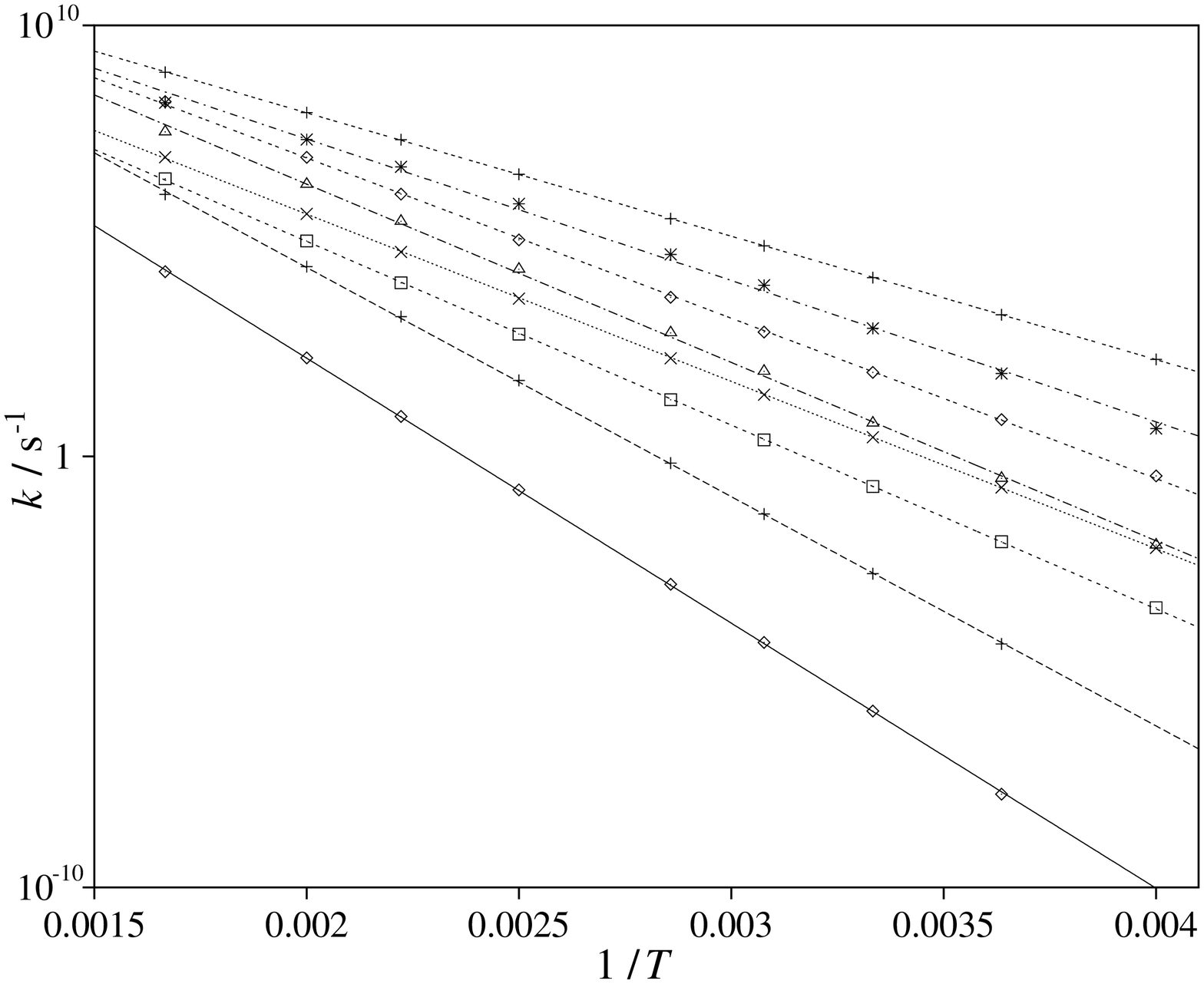,width=8.2cm}
\vglue0.1cm
\begin{minipage}{8.5cm}
\caption{\label{fig:barrier} 
Arrhenius plots for the rates of interfunnel passage. 
The data points are derived from the eigenvalues of the matrix ${\bf \tilde W}$
and the lines are fits to the form $k=A\exp(-E_a/kT)$.
In ascending order in the figure 
the lines are for the 
$5\times 5\times 3\rightarrow 8\times 3\times 3$, 
$8\times 3\times 3\rightarrow 5\times 5\times 3$, 
$5\times 5\times 3\rightarrow 6\times 4\times 3$, 
$6\times 4\times 3\rightarrow 5\times 5\times 3$, 
$6\times 4\times 3\rightarrow 5\times 4\times 4(O)$, 
$6\times 4\times 3\rightarrow 5\times 4\times 4(L)$, 
$5\times 4\times 4(O)\rightarrow 6\times 4\times 3$ and
$5\times 4\times 4(L)\rightarrow 6\times 4\times 3$ processes.  
}
\end{minipage}
\end{figure}
\end{center}

By calculating and diagonalizing  ${\bf \tilde W}$ at a series of temperatures
we can examine the temperature dependence of the interfunnel rate constants. 
For most of the interfunnel processes, the rate constants that we obtain
fit well to the Arrhenius form, $k=A\exp(-E_a/kT)$, 
where $E_a$ is the activation energy and $A$ is the prefactor (Figure \ref{fig:barrier}).

In our previous study of the interconversion mechanisms in 
(NaCl)$_{35}$Cl$^-$ we estimated the activation energies in order to compare with the experimental values. 
Using the analogy to a simple one-step reaction, we equated the activation energy
with the difference in energy between the highest-energy transition
state on the lowest-energy path between the relevant nanocrystals and
the lowest-energy minimum of the starting nanocrystal.\cite{Doye99a}
However, it was not clear how well this analogy would work. 
In Table \ref{table:barriers} we compare these estimates 
with the activation energies obtained from our master equation results.
There is good agreement, confirming the utility of the approximate approach.
Similar agreement has also been found for the interfunnel dynamics of a 
38-atom Lennard-Jones cluster.\cite{Miller99b}

A simple explanation for this correspondence can be given.
We first label the minima along the lowest-barrier path between the 
two funnels in ascending order and define $l$ so that the highest-energy 
transition state on this path lies between minima $l-1$ and $l$.
If the minima behind the highest-energy transition state (i.e.\ 1 to $l-1$) are in local equilibrium, 
then the occupation probability of the minimum $l-1$ is given by 
\begin{equation}
{p_{l-1}\over p_1}\propto \exp\left(-(E_{l-1}-E_1)/ kT\right)
\end{equation}
Then, if the rate of interfunnel flow, $k^+ p_A$, is equated to the rate of passage 
between minima $l-1$ and $l$, 
\begin{eqnarray}
k^+ p_A&\propto & p_{l-1} \exp(-(E^\dagger_{l-1,l} - E_{l-1}) / kT)\nonumber\\ 
\label{eq:barrier}
 &\propto & p_1 \exp(-(E^\dagger_{l-1,l} - E_{1}) / kT ),
\end{eqnarray}
Therefore, if the occupation probability for funnel A is dominated by 
the occupation probability for the lowest-energy minimum in the funnel, i.e.\ $p_A\approx p_1$ for all $T$,
then the activation energy is equal to 
the energy difference between the highest-energy transition state on the lowest-barrier path
and the energy of the lowest-energy minimum in the starting funnel.

We should note that in the above derivation the interfunnel probability flow is
assumed to all pass through a single transition state.
However, if there is competition between two paths,
one with a low barrier and a small prefactor and one with a larger barrier 
and a large prefactor, we expect that the low-barrier path would
dominate at low temperature and the high-barrier path at high temperature.
This behaviour would give rise to a interfunnel rate constant with a positive
curvature in an Arrhenius plot. 
However, the lines are either straight or have a small amount of negative curvature.

The lack of positive curvature, and the agreement between the estimated and 
the observed activation energies, probably indicates that the interfunnel probability flow is dominated
by paths which pass through the highest-energy transition state on the lowest barrier path.
It is interesting that, on a PES with so many minima and transition states, a single
transition state can have such a large influence on the dynamics.

At low enough temperature, $d(p_1/p_A)/dT \approx 0$, 
and so the interfunnel barrier height can be measured with respect to 
the lowest-energy minimum in the starting funnel (Equation (\ref{eq:barrier})). 
However, as the occupation probabilities of other minima in the funnel becomes significant 
relative to that for the lowest-energy minimum in the funnel, the ratio $p_1/p_A$ decreases,
thus giving the lines in the Arrhenius plot their slight negative curvature.
In other words, the apparent activation energy decreases with increasing temperature, because
the barrier height should be measured with respect to some kind of average minimum energy for the funnel,
perhaps $E_A=\sum_{i\in A} p_i E_i$.
The negative curvature is most pronounced when the occupation probabilities in a funnel change considerably.
For example, the $5\times 4\times 4 (L) \rightarrow 6\times 4\times 3$ 
rate constant has the most curvature because minimum L has a particularly low 
vibrational entropy leading to population of other 
minima within that funnel (Figure \ref{fig:disconnect}).

In Table \ref{table:barriers} we also compare our activation energies to the two
experimental values. Our values are too large by 0.24 and 0.49$\,$eV. 
There are a number of possible sources of error. 
Firstly, the samples of minima and transition state provide only 
an incomplete characterization of the PES and 
so it is possible that the nanocrystals are connected by 
undiscovered lower-barrier paths. However, we believe that our sample
of minima is a good representation of the low-energy regions of the PES and consider 
it improbable that undiscovered pathways could account for all of the discrepancy.
Secondly, in our calculations the input rate constants $k^\dagger_{ij}$ were 
calculated on the basis of the harmonic approximation. 
Although this is likely to have a significant effect on the absolute values
of the interfunnel rate constants and prefactors (the latter are too large 
compared to the experiment values\cite{Hudgins97a}), it should not have such a significant effect on 
the activation energies. Instead, we consider the most likely source of the
discrepancy between theory and experiment to be inaccuracies in the potential. 
When polarization is included by using the Welch potential, the estimated barriers
become closer to the experimental values (the discrepancies are then 0.16 and 0.34$\,$eV)
but significant differences still remain.\cite{Doye99a}
For better agreement we may need a potential that allows the properties of the 
ions to depend on the local environment. Unfortunately, although 
such potentials have been developed for a number of systems,\cite{Wilson96a,Wilson96b} 
one does not yet exist for sodium chloride.

\section{Conclusion}

The (NaCl)$_{35}$Cl$^-$ potential energy surface has a multiple-funnel topography. 
Structurally, the different funnels correspond to rocksalt 
nanocrystals with different cuboidal forms. The large potential energy barriers
between the funnels causes the time scales for escape from metastable nanocrystals to 
be far longer than those accessible by conventional dynamics simulations.
Therefore, we examined the interfunnel dynamics by applying the master equation approach to a
database of (NaCl)$_{35}$Cl$^-$ minima and transition states.
The slowest rate constant we obtained was 
$k_{5\times 5\times 3\rightarrow 8\times 3\times 3}=1.46 \times 10^{-8}\,{\rm s}^{-1}$ at $T=275$K.

Using a net flow index we were able to identify the eigenvalues of the transition matrix, ${\bf W}$,
which correspond to interfunnel probability flow.
Thus, we were able to obtain rate constants and activation energies for the interconversion
of the different nanocrystals. 
One particularly interesting finding is that the activation energies 
correspond fairly closely to the potential energy differences between 
the highest-energy transition state on the lowest-energy path between two nanocrystals
and the lowest-energy minimum of the starting nanocrystal.
This is the result one might expect by a simple extrapolation from the dynamics 
of a simple molecular reaction. 
However, it holds
despite the multi-step, and potentially multi-path, nature of the interfunnel dynamics.
The question of whether this result is generally true for interfunnel dynamics involving large 
potential energy barriers
or reflects some of the particulars of the (NaCl)$_{35}$Cl$^-$ system is 
an interesting subject for further research.
We already know that this simplification holds for the 38-atom Lennard-Jones cluster.\cite{Miller99b}

\acknowledgements

J.P.K.D. is the Sir Alan Wilson Research Fellow at Emmanuel College, Cambridge.
D.J.W.\ is grateful to the Royal Society for financial support. 
We would like to thank Mark Miller for helpful discussions.

\begin{table}
\begin{center}
\begin{minipage}{8.5cm}
\caption{\label{table:barriers}
Activation energies, $E_a$, and prefactors, $A$, for the interfunnel rate constants 
obtained from the fits in Figure \ref{fig:barrier}. 
The activation energies are compared to 
$\Delta E$, the difference in energy between the highest-energy transition
state on the lowest-energy path directly between the two nanocrystals and
the lowest-energy minimum of the starting nanocrystal,\cite{Doye99a}
and experimental values for some of the processes.\cite{Hudgins97a}
As probability flow out of minimum L leads to population of both the 
$6\times 4\times 3$ and $5\times 5\times 3$ nanocrystals we give $\Delta E$ for both pathways.
}
\vglue0.1mm
\begin{tabular}{llcccc}
 & & \multicolumn{3}{c}{barrier/eV} \\
\cline{3-5}
 \hfil From & \hfil To & 
$E_a$ & $\Delta E$ & Expt. & $A$/s$^{-1}$ \\
\hline
\noalign{\vspace{1pt}}
 $6\times 4\times 3$ & $5\times 5\times 3$ & 0.770 & 0.776 & $0.53\pm 0.05$ & $1.43 \times 10^{13}$ \\
 $5\times 5\times 3$ & $6\times 4\times 3$ & 0.846 & 0.840  & & $2.26 \times 10^{13}$ \\
 $8\times 3\times 3$ & $5\times 5\times 3$ & 1.055 & 1.055 & $0.57\pm 0.05$ & $1.05 \times 10^{15}$ \\
 $5\times 5\times 3$ & $8\times 3\times 3$ & 1.220 & 1.211 & & $3.77 \times 10^{14}$ \\
 $5\times 4\times 4$ L & $6\times 4\times 3$ & 0.651 & 0.668 & & $8.46 \times 10^{13}$ \\
 $5\times 4\times 4$ L & $5\times 5\times 3$ & & 0.618 & & \\
 $6\times 4\times 3$ & $5\times 4\times 4$ L & 0.822 & 0.810 & & $3.95 \times 10^{14}$ \\
 $5\times 5\times 3$ & $5\times 4\times 4$ L & & 0.823 & & \\
 $5\times 4\times 4$ O & $6\times 4\times 3$ & 0.568 & 0.560 & & $4.97 \times 10^{13}$ \\
 $6\times 4\times 3$ & $5\times 4\times 5$ O & 0.738 & 0.738 & & $2.32 \times 10^{14}$ \\
\end{tabular}
\end{minipage}
\end{center}
\end{table}

\end{multicols}
\end{document}